\documentclass[runningheads]{svmult}

\usepackage{makeidx}   
\usepackage{graphicx}  
\usepackage{subeqnar}  
\usepackage{multicol}  
\usepackage{physprbb}  
\makeindex             

\newcommand{\magcir}{\ \raise -2.truept\hbox{\rlap{\hbox{$\sim$}}\raise5.truept
 	\hbox{$>$}\ }}	
\newcommand{\mincir}{\ \raise -2.truept\hbox{\rlap{\hbox{$\sim$}}\raise5.truept
        \hbox{$<$}\ }}

\begin{document}
\title*{High-redshift QSOs in the GOODS}
\toctitle{High-redshift QSOs in the GOODS}
%
%
\titlerunning{High-$z$ QSOs in the GOODS}
%
\author{S.Cristiani\inst{1}
\and D.M.Alexander\inst{2,10}
\and F.Bauer\inst{2}
\and W.N.Brandt\inst{2}
\and E.T.Chatzichristou\inst{3}
\and F.Fontanot\inst{4}
\and A.Grazian\inst{5}
\and A.Koekemoer\inst{6}
\and R.A.Lucas\inst{6}
\and J.Mao\inst{11}
\and P.Monaco\inst{4}
\and M.Nonino\inst{1}
\and P.Padovani\inst{6,9}
\and D.Stern\inst{7}
\and P.Tozzi\inst{1}
\and E.Treister\inst{3}
\and C.M.Urry\inst{3}
\and E.Vanzella\inst{8}
}
\authorrunning{Stefano Cristiani et al.}
%
%
\institute{INAF-Osservatorio Astronomico, Via Tiepolo 11, I-34131 Trieste, Italy
\and Department of Astronomy and Astrophysics, Pennsylvania State
University, 525 Davey Lab, University Park, PA 16802
\and Department of Astronomy, Yale University, PO Box
208101, New Haven, CT06520
\and Dipartimento di Astronomia dell'Universit\`a, Via Tiepolo
11, I-34131 Trieste, Italy
\and INAF-Osservatorio Astronomico di Roma, via Frascati 33, I-00040
Monteporzio, Italy
\and Space Telescope Science Institute, 3700 San Martin
Dr., Baltimore, MD 21218
\and Jet Propulsion Laboratory, California Institute of Technology, 
Mail Stop 169-506, Pasadena, CA 91109
\and European Southern Observatory,
K.Schwarzschild-Stra\ss e 2, D-85748 Garching, Germany
\and ESA Space Telescope Division
\and Institute of Astronomy, Madingley Road, Cambridge, CB3 0HA, UK
\and SISSA, via Beirut 4, I-34014 Trieste, Italy
}
\maketitle              



\section{Introduction}
QSOs are intrinsically luminous and therefore can be seen rather
easily at large distances; but they are rare, and finding them
requires surveys over large areas. As a consequence, at present, the
number density of QSOs at high redshift is not well known.  Recently,
the Sloan Digital Sky Survey (SDSS) has produced a breakthrough,
discovering QSOs up to $z=6.43$ \cite{Fan03} and building a sample of
six QSOs with $z>5.7$.  The SDSS, however, has provided
information only about very luminous QSOs ($M_{1450} \mincir -26.5$),
leaving unconstrained the faint end of the high-$z$ QSO Luminosity
Function (LF), which is particularly important to understand the
interplay between the formation of galaxies and super-massive black
holes (SMBH) and to measure the QSO contribution to the UV ionizing
background \cite{Madau99}.
New deep multi-wavelength surveys like the Great Observatories Origins
Deep Survey (GOODS), described by M.Dickinson, M.Giavalisco and
H.Ferguson in this conference, provide significant constraints on the space
density of less luminous QSOs at high redshift. 
Here we present a search for high-$z$ QSOs, identified in the two
GOODS fields on the basis of deep imaging in the optical (with {\em
HST}) and X-ray (with {\it Chandra}), and discuss the allowed space
density of QSOs in the early universe, updating the results presented in
\cite{cristiani04}.

\section{The Database}
The optical data used in this search have been obtained with the ACS onboard
HST, as described by M.Giavalisco.
Mosaics have been created from the first three epochs of
observations, out of a total of five, in the bands
$F435W(B_{435})$, $F606W(V_{606})$, $F775W (i_{775})$,
$F850LP(z_{850})$. The catalogs used to select high-$z$ QSOs have
been generated using the SExtractor software, performing the
detection in the $z_{850}$ band and then using the isophotes
defined during this process as apertures for photometry in the
other bands. This is a common practice avoiding biases due to
aperture mismatch coming from independent detections.

The X-ray observations of the HDF-N and CDF-S consist of 2 Ms and
1 Ms exposures, respectively, providing the deepest views of the Universe in
the 0.5--8.0~keV band. The X-ray completeness limits over $~90\%$
of the area of the GOODS fields are similar, with flux limits
(S/N$=5$) of $~1.7\times10^{-16}$~erg~cm$^{-2}$~s$^{-1}$
(0.5--2.0~keV) and $~1.2 \times10^{-15}$ ~erg~cm$^{-2}$~s$^{-1}$
(2--8~keV) in the HDF-N field, and $\approx
2.2\times10^{-16}$~erg~cm$^{-2}$~s$^{-1}$ (0.5--2.0~keV) and
$\approx 1.5\times10^{-15}$~erg~cm$^{-2}$~s$^{-1}$ (2--8~keV) in
the CDF-S field. The sensitivity at the aim point is about 2 and 4
times better for the CDF-S and HDF-N, respectively. As an example,
assuming an X-ray spectral slope of $\Gamma=$~2.0, a source
detected with a flux of $1.0\times10^{-16}$~erg~cm$^{-2}$~s$^{-1}$
would have both observed and rest-frame luminosities of  $8\times
10^{42}$~erg~s$^{-1}$, and $3\times 10^{43}$~erg~s$^{-1}$ at
$z=3$, and $z=5$, respectively (assuming no Galactic absorption).
\cite{Alexander03} produced point-source catalogs for the HDF-N and
CDF-S and \cite{Giacconi02}
for the CDF-S with more than 1000 detected sources in total.

\section{The Selection of the QSO candidates}
We have carried out the selection of the QSO candidates in
the magnitude interval $22.45 < z_{850} < 25.25$. Four optical
criteria have been tailored on the basis of typical QSO SEDs
in order to select QSOs at progressively higher redshift in the
interval $3.5 \mincir z \mincir 5.2$.  The criteria have been applied
independently and produced in total $645$ candidates in the CDF-S
and $557$ in the HDF-N. They select a broad range of high-$z$ AGN, not
limited to broad-lined (type-1) QSOs, and are less stringent than
those typically used to identify high-$z$ galaxies. 
We have checked the criteria against QSOs and
galaxies known in the literature within the magnitude and redshift
ranges of interest confirming the high completeness of the adopted
criteria.

Below $z \simeq 3.5$ the typical QSO colors in the ACS bands move
close to the locus of stars and low-redshift galaxies. Beyond $z
\simeq 5.2$ the $i-z$ color starts increasing and infrared bands
would be needed to identify QSOs efficiently with an ``$i$-dropout''
technique.

\section{Match with Chandra Sources}
The optical candidates  have been matched with X-ray sources
detected by Chandra within an error radius corresponding to the
$3~\sigma$ X-ray positional uncertainty. With this tolerance the
expected number of false matches is five and indeed two
misidentifications, i.e. cases in which a brighter optical source
lies closer to the X-ray position, have been rejected (both in the
CDF-S).

The sample has been reduced in this way to 11 objects in the CDF-S
and 6 in the HDF-N. Type-1 QSOs with $M_{1450}<-21$, given the
measured dispersion in their optical-to-X-ray flux ratio 
\cite{Vignali03}, are  detectable in our X-ray observation up to $z
\magcir 5.2$. Conversely, any $z>3.5$ source in the GOODS region
detected in the X-rays must harbor an AGN  ($L_x(0.5-2~{\rm keV})
\magcir 10^{43}~{\rm erg~s^{-1}}$).

\section{Redshifts of the QSO candidates}
Twelve objects out of the 17 selected have spectroscopic
confirmations. Nine are QSOs with redshifts between $2.6$ and $5.2$.
Three are reported to be galaxies, and the relatively large offsets
between the X-ray and optical positions suggest that they could be
misidentifications. Both quasars of Type I and II are detected.

Photometric redshifts of the 17 QSO candidates have been estimated by
comparing with a $\chi^2$ technique (see \cite{Arnouts99} for details)
the observed ACS colors to those expected on the basis of {\it a)} 
the typical QSO SEDs; {\it b)} a library of template SEDs of galaxies 
(the ``extended Coleman'' of \cite{Arnouts99}).
For the nine QSOs with spectroscopic confirmation the photometric 
redshifts are in good agreement with the observed ones. 
In general the estimates {\em a)} and {\em b)} are similar,
since the color selection criteria both for galaxies and QSOs rely on
a strong flux decrement in the blue part of the spectrum - due to the
IGM and possibly an intrinsic Lyman limit absorption - superimposed on
an otherwise blue continuum. If we limit ourselves to the redshift
range $z>4$ - where the selection criteria and the photometric redshifts
are expected to be most complete and reliable - in addition to one
spectroscopically confirmed QSO (HDF-N 123647.9+620941, $z=5.186$) we
estimate that between 1-2 more QSOs are present in the GOODS,
depending on whether galaxy or QSO SEDs are adopted for the
photometric redshifts. This brings the total to no more than three QSOs
with redshift larger than 4.
\begin{figure}[ht]
\begin{center}
\includegraphics[width=1\textwidth]{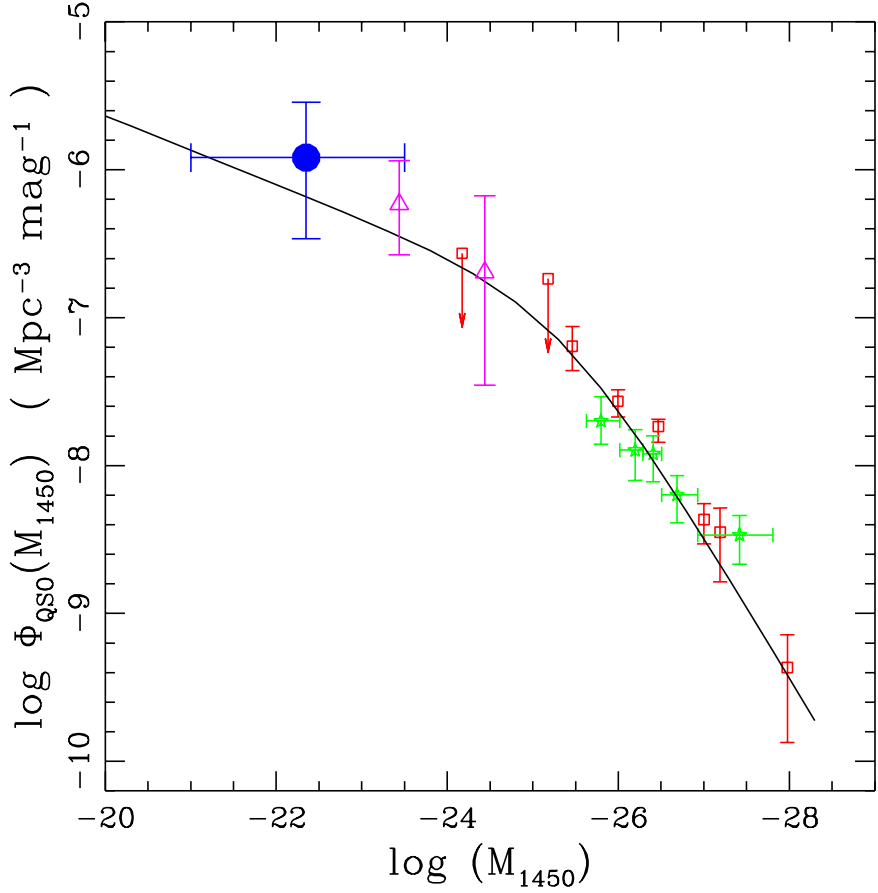}
\end{center}
\caption[]{
Differential QSO luminosity function 
at $z\sim4-5$.  Squares: data from \cite{Kennefick95}
and upper limits from \cite{Kennefick96} at $z\sim4.4$.  Stars: SDSS
\cite{fan01b} computed at $3.5<z<5$ assuming a power-law decline
and rescaled to $z=4.3$.  Triangles: COMBO-17 LF \cite{COMBO17}, for
$4.2<z<4.8$.  Circle: GOODS LF at $4<z<5.2$.
The continuous line shows the prediction of the PDE model described in
the text. 
All LFs have been rescaled to a cosmology with $h, \Omega_{\rm tot},
\Omega_m, \Omega_{\Lambda} = 0.7, 1.0, 0.3, 0.7$.}
\label{fig1}
\end{figure}
\section{Comparison with the models}
Let us compare now the QSO counts observed in the $z_{850}$ band with
two phenomenological and two more physically motivated models. The
double power-law fit of the 2QZ QSO LF \cite{boyle00} has
been extrapolated for $z>2.7$ (the peak of QSO activity) in a way to
produce a power-law decrease of the number of bright QSOs by a
factor 3.5 per unit redshift interval, consistent with the
$3.0^{+1.3}_{-0.9}$ factor found by \cite{fan01b} for the bright
part of the LF. The extrapolation is carried out either as a Pure
Luminosity Evolution (PLE) or a Pure Density Evolution (PDE). The
PLE model predicts about 17 QSOs with $z_{850}<25.25$ at redshift
$z>4$ (27 at $z>3.5$) in the 320 arcmin$^2$ of the two GOODS fields, and
is inconsistent with the observations at a more than $3 \sigma$
level. 
The PDE estimate is 2.9 QSOs at $z>4$ (6.7 at $z>3.5$).

It is possible to connect QSOs with dark matter halos (DMH) 
formed in hierarchical
cosmologies with a minimal set of assumptions (MIN model, e.g.
\cite{haiman01}): {\em a)} QSOs are hosted in newly formed halos
with {\em b)} a constant SMBH/DMH mass ratio $\epsilon$ and {\em
c)} accretion at the Eddington rate. The bolometric LF of QSOs is
then expressed by:
\begin{equation} 
\Phi(L|z)dL =  n_{\rm PS}(M_H|z) \int_{t(z)-t_{\rm duty}}^{t(z)}
\hskip -11mm
P(t_f|M_H,t(z))dt_f \epsilon^{-1} {{dL}\over{L_{\rm Edd}}} 
\label{eq:hh}
\end{equation}
where the abundance of DMHs of
mass $M_H$ is computed using the Press \& Schechter \cite{ps74} recipe,
the distribution $P(t_f|M_H,t(z))dt_f$ of the formation times
$t_f$ follows \cite{lc93}, $t_{\rm duty}$ is the QSO duty cycle
and $L_{Edd}=10^{4.53}~L_{\odot}$ is the Eddington luminosity of a
$1\ M_\odot$ SMBH. This model is known to overproduce the number
of high-$z$ QSOs \cite{haiman99} and in the present case predicts
$151$ QSOs at $z>4$ ($189$ at $z>3.5$).

To cure the problems of the MIN model feedback effects have been
invoked, for example assuming that the QSOs shine a $t_{\rm delay}$ time
after DMH formation (DEL model; \cite{Monaco00}, \cite{granato01}. 
The QSO LF is then computed at a
redshift $z'$ corresponding to $t(z)-t_{\rm delay}$.  The
predictions of the DEL model are very close to the PDE, with $3.2$
QSOs expected at $z>4$ ($4.8$ at $z>3.5$), in agreement with the
observations.
While all the models presented here are consistent with the recent
QSO LF measurement of the COMBO-17 survey \cite{COMBO17}
at $z \simeq 2-3$, only the
PDE and DEL models fit the COMBO-17 LF in the range $4.2<z<4.8$ and
$M_{1450}<-26$.
\section{Conclusions}
At $z>4$ the space density of moderate luminosity ($M_{1450} \simeq
-23$) QSOs is significantly lower than the prediction of simple
recipes matched to the SDSS data, such as a PLE evolution of the LF or
a constant universal efficiency in the formation of SMBH in DMH. 
A flattening of the observed high-$z$ LF  is required
below the typical luminosity regime ($M_{1450} \mincir -26.5$)
probed by the SDSS. 
An independent indication that this flattening
must occur comes from the statistics of bright lensed QSOs
observed in the SDSS \cite{Wyithe02} that  would be much larger if
the LF remains steep in the faint end. 
A similar result has been obtained at $5 \mincir z \mincir 6.5$, by
\cite{Barger03}. The QSO contribution to the UV background is
insufficient to ionize the IGM at these redshifts. 
This is an indication that
at these early epochs the formation or the feeding of SMBH is
strongly suppressed in relatively low-mass DMH, as a consequence
of feedback from star formation \cite{granato01} and/or
photoionization heating of the gas by the UV background 
\cite{haiman99}, accomplishing 
a kind of inverse hierarchical scenario.

\end{document}